\documentclass[aps,prm,twocolumn,superscriptaddress,notitlepage]{revtex4-2}
\usepackage{graphicx}
\usepackage{dcolumn}
\usepackage{bm}
\usepackage{amsmath,amssymb,amsthm,graphicx, color}
\usepackage{times}
\usepackage{dsfont}
\usepackage{braket}
\usepackage{multirow}
\usepackage[colorlinks,allcolors=blue]{hyperref}
\usepackage[utf8]{inputenc}
\usepackage{dcolumn}
\usepackage{bm}
\usepackage{hyperref}
\usepackage{times}
\usepackage{braket}
\usepackage{float}

\begin{document}

\title{Robust quantized thermal conductance of Majorana floating edge bands in d-wave superconductors}

\author{Yanmiao Han}
\altaffiliation{These authors contributed equally to this work.}
\affiliation{School of Physics, Beihang University, Beijing 102206, China}

\author{Yu-Hao Wan}
\altaffiliation{These authors contributed equally to this work.}
\email[Contact author:] {wanyh@stu.pku.edu.cn}

\affiliation{International Center for Quantum Materials and School of Physics, Peking University, Beijing 100871, China}

\author{Zhaoqin Cao}
\affiliation{School of Mathematical Sciences, Universiti Sains Malaysia, Penang 11800, Malaysia}

\author{Rundong Zhao}
\email[Contact author:] {rdzhao@buaa.edu.cn}
\affiliation{School of Physics, Beihang University, Beijing 102206, China}

\author{Qing-Feng Sun}
\email[Contact author:] {sunqf@pku.edu.cn}
\affiliation{International Center for Quantum Materials and School of Physics, Peking University, Beijing 100871, China}
\affiliation{Hefei National Laboratory, Hefei 230088, China}

\begin{abstract}
We propose and characterize a new class of Majorana boundary states, i.e., floating Majorana edge bands (FMEBs), which emerge in two-dimensional (2D) superconductors that break time-reversal symmetry yet host helical-like transport. In contrast to conventional chiral or helical edge modes, FMEBs form isolated, momentum-separated counterpropagating Majorana modes detached from the bulk continuum. We identify a minimal mechanism for their emergence via anisotropic Wilson masses in a two-band Bogoliubov–de Gennes (BdG) model, and demonstrate their microscopic realization in a quantum anomalous Hall (QAH) insulator proximitized by a $d$-wave superconductor. Using nonequilibrium Green’s function (NEGF) simulations, we uncover clear transport fingerprints: a quantized total thermal conductance in two-terminal devices, and a robust half-quantized plateau in four-terminal geometries that cleanly distinguishes FMEBs from chiral $\mathcal{N}= \pm 2$ QAH phases. This thermal response remains remarkably stable under finite temperature, moderate long-range disorder, and finite chemical potential. Our findings establish FMEBs as an experimentally accessible route toward helical-like Majorana transport in systems without time-reversal symmetry, with direct implications for topological quantum computation.
\end{abstract}

\maketitle

\section{Introduction}

Topological superconductors host Majorana quasiparticles, which are real fermionic modes equal to their own antiparticles, and give rise to non-Abelian boundary excitations\cite{frolov2020topological,sato2017topological, PhysRevApplied.16.014024}. In one dimension, unpaired Majorana end states appear in topological superconducting wires, while in two dimensions, a chiral $p$-wave superconductor supports both a single chiral Majorana edge channel and Majorana zero modes in vortices\cite{PhysRevLett.86.268}. These paradigms were established by Read and Green’s analysis of parity- and time-reversal-breaking paired states and by Kitaev’s exactly solvable 1D chain , and they underpin the vision of fault-tolerant quantum computation based on non-Abelian braiding\cite{ReadGreen2000,Kitaev2001,Alicea2012,Beenakker2013ARCMP,han_impact_2025,Prada2020NatRevPhys}.

Various systems have been predicted to host propagating helical\cite{PhysRevLett.108.036803, PhysRevLett.122.187001} or chiral\cite{PhysRevB.61.10267,PhysRevLett.121.256801} Majorana modes with or without time-reversal symmetry, respectively.
A powerful and experimentally versatile route to Majorana modes is topological proximity superconductivity\cite{LiChanYao2015PRB, Zareapour2012NatCommun, wang2012coexistence, Zhao2018PRB}.
Theoretical models have shown that coupling a conventional $s$-wave superconductor to topological electronic states can generate distinct types of topological superconductivity. When deposited on a 3D topological-insulator surface, the proximity effect can yield a time-reversal-invariant topological superconductor hosting Majorana bound states. In spin-orbit-coupled semiconductor nanowires subjected to a Zeeman field, it produces 1D topological superconductivity, while in quantum anomalous Hall (QAH) systems proximitized by an $s$-wave superconductor, it gives rise to a chiral topological phase\cite{FuKane2008PRL,Lutchyn2010,Oreg2010,QiHughesZhang2010,addQAH}.
Experimentally, proximity-induced superconductivity has been realized on quantum-spin-Hall edges in HgTe and InAs/GaSb quantum wells, enabling gate-tunable 1D topological Josephson junctions\cite{Hart2014,Pribiag2015}.
More recently, QAH-superconductor heterostructures have emerged as a promising platform for chiral Majorana modes, providing a solid-state realization of a single Majorana edge channel along a magnetic topological boundary\cite{He2017Science_Retracted, Kayyalha2020Science, Huang2018PRB, PhysRevLett.120.107002, Lian2018PNAS, Shen2020PNAS, Banerjee2018Nature, Yokoi2021Science, Simon2018PRB}.

Conventionally, edge bands connect the bulk conduction and valence bands and exist only within a restricted region of momentum space.
Recent studies have revealed a distinct scenario in which edge dispersions can extend across the entire Brillouin zone and even detach from the bulk continua, forming floating edge bands (FEBs) inside the band gap~\cite{Topp2017PRX,Zhu2018NatCommun,Ma2024PhotonRes}.
For example, in electronic systems FEBs\cite{LiZhang2025PRB} can be engineered by introducing additional altermagnetic\cite{Smejkal2022PRX,Krempasky2024Nature,Lee2024PRL} and out-of-plane Zeeman fields. This unconventional boundary phenomenon has been intensively explored not only in condensed-matter settings such as magnetic and nonsymmorphic materials, but also in photonic, acoustic, and other synthetic platforms~\cite{Altland2024PRX,Nakamura2024arXiv,ma2024asymmetric}.
Moreover, FEBs are closely related to other intriguing topological phases, including higher-order topological states~\cite{PhysRevB.107.174101,PhysRevB.106.155144}.

Despite the progress achieved in electronic and photonic systems, it remains an open question whether an intrinsic superconducting analogue of FEB exists, how such a state could emerge microscopically, and what transport signatures would distinguish it from conventional chiral or helical Majorana edges.

In this work, we introduce the concept of floating Majorana edge bands (FMEBs) and explore their physical origin, microscopic realization, and transport signatures. Starting from a minimal two-band BdG lattice model, we show that anisotropy in the Wilson mass terms drives a transition from a conventional chiral topological superconductor with Chern number $\mathcal{N}=1$ to a gapless edge phase with $\mathcal{N}=0$, where the boundary reconstructs into momentum-separated, counterpropagating Majorana modes. We then demonstrate that such anisotropy arises naturally in a QAH insulator coupled to a $d$-wave superconductor, where the pairing form factor renormalizes the two Majorana blocks in opposite directions. Using the nonequilibrium Green’s function (NEGF) method, we reveal distinct thermal transport fingerprints of FMEBs: two-terminal conductance remains quantized, while a four-terminal, single-edge setup reveals a robust half-quantized thermal plateau. We further establish the stability of this transport signature against finite temperature, long-range correlated disorder, and moderate chemical potential, supporting the experimental feasibility of detecting FMEBs. The results establish FMEBs as a distinct class of superconducting boundary states hosting helical-like Majorana modes stabilized by anisotropy, and reveal their characteristic half-quantized thermal response, thereby extending the landscape of Majorana transport phenomena.

The remainder of this paper is organized as follows.
In Sec.\ref{sec2}, we introduce the minimal two-band BdG model and show how anisotropic Wilson mass leads to the FMEB phase.
In Sec.\ref{sec3}, we demonstrate that this mechanism naturally arises in a QAH insulator proximitized by a $d$-wave superconductor.
In Sec.\ref{sec4}, we analyze the corresponding transport characteristics using the NEGF method, identifying clear quantized and half-quantized thermal conductance signatures that distinguish FMEBs from conventional QAH phases.
Sec.\ref{sec5} examines the robustness of these transport features against temperature, disorder, and chemical potential.
Finally, Sec.\ref{sec6} summarizes the main findings.

\section{Minimal two-band model for FMEBs}\label{sec2}
\begin{figure}[htbp]
    \includegraphics[width=1\linewidth]{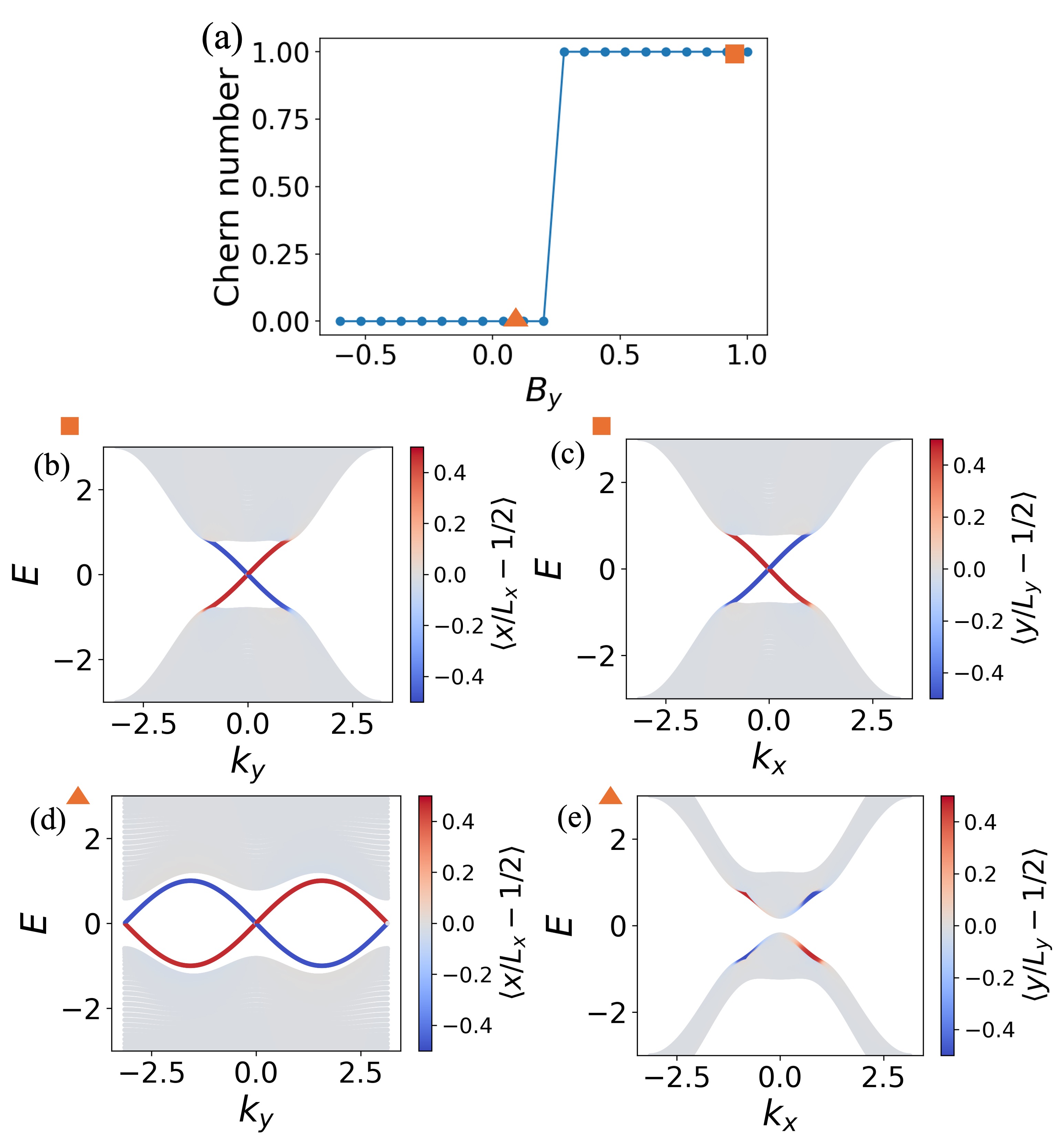}
        \caption{
    Bulk and edge spectra of the anisotropic BdG-QWZ model, illustrating the emergence of FMEBs driven by Wilson mass anisotropy. Here, the model parameters are fixed as \(m=1\), \(B_x=1\), while \(B_y\) is varied to induce the topological transition.
    (a) Chern number computed from the momentum-space Hamiltonian as a function of \(B_y\).
    (b,c) Ribbon spectra at the isotropic point \(B_y=1\) for \(x\)-open and \(y\)-open geometries (with ribbon width \(L_x=40\) and \(L_y=40\), respectively), both exhibiting a single chiral Majorana edge mode traversing the bulk gap.
    (d,e) For anisotropic parameters \(B_y=0.1\), the \(x\)-open ribbon hosts a FMEB largely detached from the bulk continuum, whereas the \(y\)-open ribbon is fully gapped.
    Color in panels (b–e) indicates the normalized transverse coordinate $\langle y/L_y - 1/2\rangle$ or $\langle x/L_x - 1/2\rangle$, to distinguish edge-localized states.}
        \label{fig1}
\end{figure}

To gain insight into the emergence of FMEB, we start from a minimal lattice model. This simplified setting allows us to clearly identify the mechanism by which the boundary reorganizes from a conventional chiral topological superconductor into a helical floating phase.
We consider a two-band BdG Hamiltonian on a square lattice, with Nambu basis $\Psi_{\mathbf k}=(c_{\mathbf k},c^\dagger_{-\mathbf k})^{T}$\cite{PhysRevB.92.064520, PhysRevB.95.245433, PhysRevB.98.045141}:
\begin{equation}
H(\mathbf k)
= v \sin k_x\,\tau_x + v \sin k_y\,\tau_y
+ M(\mathbf k)\tau_z,
\label{eq:H_lattice}
\end{equation}
where $\tau_{x,y,z}$ act in the electron/hole pseudospin and $v$ sets the Fermi velocity. The topology of the system is governed by the mass term
\begin{equation}
M(\mathbf k) = m - 2B_x - 2B_y + 2B_x \cos k_x + 2B_y \cos k_y ,
\label{eq:Mk}
\end{equation}
with $B_{x,y}$ being the (generally anisotropic) Wilson mass coefficients along the $x$ and $y$ directions.
The Chern number can be calculated using
\begin{equation}\label{1}
    \mathcal{C} = \frac{1}{2\pi} \sum_n \int_{\text{BZ}} d^2k \ \Omega_n(\mathbf{k}),
\end{equation}
where $\Omega_n(\mathbf{k})$ is momentum-dependent Berry curvature for the $n$th band:
\begin{equation}\label{2}
    \Omega_n(\mathbf{k}) = - \sum_{n' \neq n} \frac{2 \, \mathrm{Im} \left[ \langle \psi_{n\mathbf{k}} | v_x | \psi_{n'\mathbf{k}} \rangle \langle \psi_{n'\mathbf{k}} | v_y | \psi_{n\mathbf{k}} \rangle \right]}{(\epsilon_{n'\mathbf{k}} - \epsilon_{n\mathbf{k}})^2},
\end{equation}
with summation over all occupied bands. Here, $|\psi_{n\mathbf{k}}\rangle$ denotes the Bloch state of the $n$th band, $v_x$ and $v_y$ are the velocity operators in the $x$ and $y$ directions, and $\epsilon_{n\mathbf{k}}$ is the energy eigenvalue of the $n$th band at momentum $\mathbf{k}$.
For parameters $m=1$ and $B_x=B_y=1$, the model realizes a topological superconducting phase with $\mathcal{C}=1$, hosting a single chiral Majorana edge mode.
Figs.~\hyperref[fig1]{1(b)} and \hyperref[fig1]{1(c)} show the corresponding ribbon spectra for $x$-open (good quantum number $k_y$) and $y$-open (good quantum number $k_x$) geometries, respectively—both exhibiting one chiral edge branch traversing the bulk gap, consistent with the bulk-boundary correspondence. We now examine how anisotropy in the Wilson mass modifies the topological character of the system.
By varying $B_y$ while fixing $m=1$ and $B_x=1$, we obtain the phase evolution summarized in Fig.\hyperref[fig1]{1(a)}.
Once $B_y$ deviates from $B_x$, the lattice loses isotropy and the system undergoes a topological phase transition.
At the isotropic point $B_x=B_y$, the model belongs to the topological superconductor phase with $\mathcal{C}=1$.
Increasing anisotropy eventually drives the bulk Chern number to zero ($\mathcal{C}=0$), signaling the collapse of the global topological invariant.
A natural question then arises: although the bulk Chern number vanishes, does the system truly become topologically trivial?
To address this, we examine the ribbon spectra. As shown in Fig.\hyperref[fig1]{1(d)}, the $x$-open ribbon exhibits a floating edge band within the bulk gap, whereas the $y$-open ribbon in Fig.\hyperref[fig1]{1(e)} is fully gapped.
The floating edge band is completely detached from the bulk continua across the entire Brillouin zone, forming an isolated band that persists within the gap without hybridizing with bulk states.
In the Majorana representation, this spectral feature corresponds to the emergence of a FMEB.

Anisotropic Wilson mass provides the minimal route to FMEBs in this model: it drives the bulk Chern number to zero while preserving an orientation-selective floating edge band.
In Sec.\ref{sec3}, we show that this anisotropy naturally arises in a QAH insulator proximitized by a $d$-wave superconductor, reproducing the same floating-band phenomenology at the microscopic level.

\section{Microscopic realization in a QAH insulator coupled to a $d$-wave superconductor}\label{sec3}

\begin{figure}[htbp]
    \centering

    \includegraphics[width=0.5\textwidth]{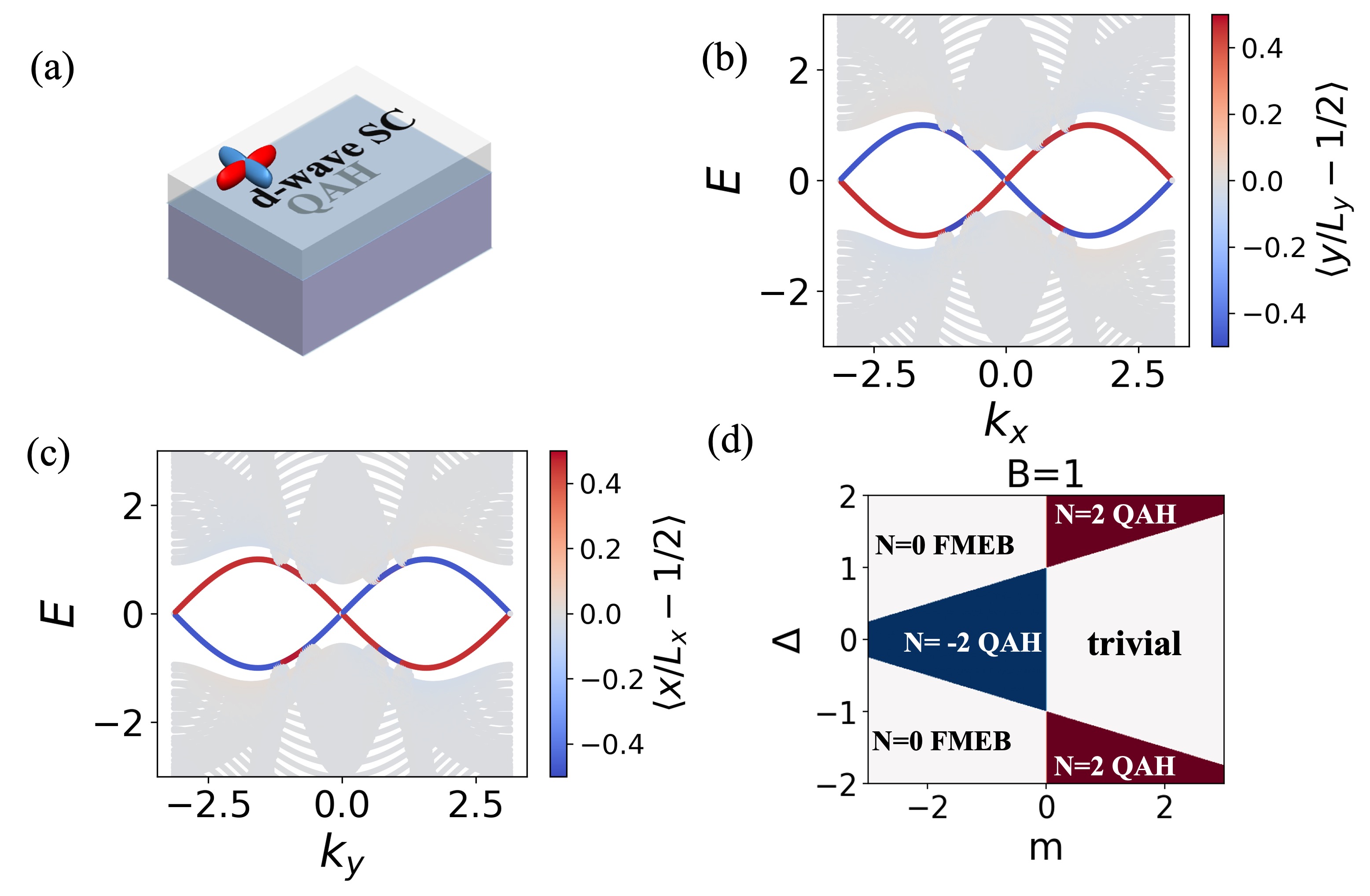}
        \caption{Realization of FMEBs in a QAH insulator proximitized by a $d$-wave superconductor. (a) Schematic illustration of the QAH system/$d$-wave superconductor heterostructure. (b,c) Ribbon spectra in the FMEB regime for $y$-open ($k_x$ good) and $x$-open ($k_y$ good) geometries, respectively. In both orientations, the edge hosts a pair of counterpropagating Majorana modes separated in momentum, forming helical-like FMEBs despite a vanishing total Chern number. Color scale encodes transverse wavefunction localization relative to the ribbon center. The spectra are calculated with ribbon width $L_x=L_y=60$, mass term $m=-1$, and $d$-wave pairing amplitude $\Delta=1.5$. (d) Phase diagram in the $(m,\Delta)$ plane at fixed Wilson mass $B=1$, showing chiral QAH phases with Chern numbers $\mathcal{N}=\pm2$ (colored regions), a trivial $\mathcal{N}=0$ wedge (white), and FMEB regimes with zero Chern number but nontrivial edge reconstruction (white pockets).}
\label{fig2}
\end{figure}

Having established that Wilson mass anisotropy ($B_x\neq B_y$) is the route to FMEB, we now turn to a microscopic platform where such anisotropy arises naturally. A minimal model for a QAH system is given by the Qi-Wu-Zhang (QWZ) Hamiltonian.
When this QAH state is coupled to a $d$ -wave superconductor, as illustrated in Fig.\hyperref[fig2]{2(a)}, the combined system realizes the desired anisotropic setting. In the Nambu basis $\Psi_{\mathbf k}=(c_{\mathbf k},\,c^\dagger_{-\mathbf k})^T$ with $c_{\mathbf k}=(c_{\uparrow,\mathbf k},\,c_{\downarrow,\mathbf k})^T$, the BdG Hamiltonian reads
\begin{equation}
H_{\rm BdG}(\mathbf k)=
\begin{pmatrix}
h_{\rm QAH}(\mathbf k) & -i\Delta(\mathbf k)\,\sigma_y \\
i\Delta(\mathbf k)\,\sigma_y & -h_{\rm QAH}^T(-\mathbf k)
\end{pmatrix},
\label{eq:BdG_4band}
\end{equation}
where $h_{\rm QAH}(\mathbf k)=
v\big(\sin k_x\,\sigma_x+\sin k_y\,\sigma_y\big)
+\Big[m+2B\big(2-\cos k_x-\cos k_y\big)\Big]\sigma_z$\cite{RevModPhys.83.1057}, $\sigma_{x,y,z}$ are Pauli matrices, $v$ is the Fermi velocity, $m$ controls magnetism, and $B$ is the Wilson mass. The proximity-induced $d$-wave pairing is taken as
\begin{equation}
\Delta(\mathbf k)=2\Delta\,(\cos k_x-\cos k_y).
\label{eq:delta_d}
\end{equation}

To convert the Nambu representation to the Majorana representation \cite{yan2021majorana}, we perform a unitary transformation:
\begin{equation}
U_M=\frac{1}{\sqrt2}
\begin{pmatrix}
1 & 0 & 0 & 1\\
0 & 1 & 1 & 0\\
0 & -1 & 1 & 0\\
-1 & 0 & 0 & 1
\end{pmatrix}.
\label{eq:UM_lat}
\end{equation}
Acting it on the BdG Hamiltonian Eq.\eqref{eq:BdG_4band} gives
\begin{equation}
\tilde{H}(\bm{k})=U_M\,H_{\rm BdG}(\bm{k})\,U_M^\dagger
=
\begin{pmatrix}
H_+(\bm{k})&0 \\0
 & H_-(\bm{k})
\end{pmatrix}.
\label{eq:H_tilde_lat}
\end{equation}
The Hamiltonian is strictly block–diagonalized, with the block reads
\begin{equation}
H_\pm(\bm{k})=v\big(\sin k_x\,\sigma_x+\sin k_y\,\sigma_y\big)+M_\pm(\bm{k})\,\sigma_z,
\label{eq:Hpm_lat}
\end{equation}
with
\begin{equation}
M_\pm(\bm{k})=m+2B\,(2-\cos k_x-\cos k_y)\ \pm\ 2\Delta(\cos k_x-\cos k_y).
\label{eq:M_lat_final}
\end{equation}
Eqs.~(\ref{eq:Hpm_lat}) and (\ref{eq:M_lat_final}) show that the $d$-wave form factor ($\cos k_x-\cos k_y$) enters the two Majorana blocks $H_{\pm}$ with opposite signs. Consequently, the pairing term renormalizes the lattice masses $M_{ \pm}(\mathbf{k})$ in opposite ways: it enhances the effective Wilson mass along one direction while reducing it along the orthogonal direction.
This corresponds to an anisotropic Wilson parameterization,
$B_x^{(+)}=B+\Delta,  B_y^{(+)}=B-\Delta, B_x^{(-)}=B-\Delta, B_y^{(-)}=B+\Delta $,
showing that the $d$-wave pairing naturally realizes the $B_x \neq B_y$ condition identified in Sec.\ref{sec2}.
In this sense, the $d$-wave term provides a microscopic route to the anisotropy responsible for FMEB.

Figs.~\hyperref[fig2]{2(b)} and \hyperref[fig2]{2(c)} display the ribbon dispersions for the $y$-open ($k_x$ conserved) and $x$-open ($k_y$ conserved) geometries, respectively. Surprisingly, FMEBs appear along both ribbon orientations. Although in Figs.~\hyperref[fig2]{2(b)} and \hyperref[fig2]{2(c)} the floating bands may visually appear to “touch” the bulk continua, this is only because the plotted spectrum overlays contributions from the two decoupled Majorana blocks; the apparent attachment involves states from different blocks, and since $H_{+}$ and $H_{-}$ are exactly block-diagonal, there is no hybridization between them.
This behavior originates from the two Majorana blocks, because the $d$-wave term enters $H_{+}$ and $H_{-}$ with opposite signs (see Eq.(\ref{eq:M_lat_final})).
Consequently, the $d$-wave pairing modifies the effective Wilson mass anisotropy in opposite directions for the two blocks: The effective mass associated with $H_{+}$ is reduced along the $x$ direction, whereas that of $H_{-}$
is reduced along the $y$ direction.
As a result, $H_{+}$ develops a floating Majorana band on the $x$-open ribbon, and $H_{-}$ produces its counterpart on the $y$-open ribbon.
Although the total bulk Chern number vanishes ( $\mathcal{N}=0$ ), each orientation supports a pair of counterpropagating Majorana channels derived from a single block, consistent with the FMEB phenomenology discussed in Sec.\ref{sec2}.




The competition between the QAH band-inversion term $m$ and the $d$-wave pairing amplitude $\Delta$ defines the overall topology of the system.
To trace this evolution, we construct the (m,$\Delta$) phase diagram (Fig.\hyperref[fig2]{2(d)}) and determine the phase boundaries analytically from the sign changes of the block masses at the four high-symmetry points\cite{asboth2016short}.
At $\Gamma=(0,0)$, $X=(\pi,0)$, $Y=(0,\pi)$, and $M=(\pi,\pi)$, one finds
\begin{equation}
\begin{aligned}
&M_\pm(\Gamma)=m,\\
&M_\pm(X)=m+4(B\pm\Delta),\\
&M_\pm(Y)=m+4(B\mp\Delta),\\
&M_\pm(M)=m+8B.
\end{aligned}
\label{eq:Mpm_TRIM}
\end{equation}
For a QWZ, the Chern number of each block is
\begin{equation}
\begin{aligned}
C_\pm=\frac12\![&\operatorname{sgn}M_\pm(\Gamma)+\operatorname{sgn}M_\pm(M)
\\&-\operatorname{sgn}M_\pm(X)-\operatorname{sgn}M_\pm(Y)].
\end{aligned}
\label{eq:Chern_sign}
\end{equation}
Because reversing the sign of $\Delta$ makes the $X$ and $Y$ points symmetry-equivalent in Eq.~(\ref{eq:Mpm_TRIM}), the two sectors acquire identical Chern numbers, $C_{-}=C_{+}$. Accordingly, the total Chern number is $N\equiv C_{+}+C_{-}=2C_{+}\in\{-2,0,+2\}$.
Phase boundaries are bulk-gap closings when any high-symmetry point's mass vanishes: $ m=0, m=-8B, m=-4(B+\Delta), m=-4(B-\Delta)$. For the parameters used in Fig.\hyperref[fig2]{2(b)} ($B=1$), these conditions yield
$m=0,\; m=-8,\; m=-4(1+\Delta),$ and $m=-4(1-\Delta)$.

Fig.\hyperref[fig2]{2(d)} shows the analytic $(m,\Delta)$ phase diagram at fixed $B=1$.
The two oblique lines $m=-4(1\pm\Delta)$ bound a trivial wedge with total Chern number $\mathcal{N}=0$ (white region).
Outside this wedge: for $m>0$ one obtains a chiral phase with $\mathcal{N}=+2$, while for $m<0$ and small $|\Delta|$, a chiral phase with $\mathcal{N}=-2$ appears.
Away from $\Delta=0$ on the $m<0$ side, two $N=0$ regions (FMEB) appear above and below the $\mathcal{N}=-2$ region.
Although their bulk Chern number vanishes, these phases are distinct from the trivial ones: the $d$-wave-induced anisotropy enforces a weak-topological condition along a specific ribbon orientation, giving rise to FMEBs, which can be unambiguously distinguished from the trivial $N=0$ phase by the nonzero winding pair $(W_0,W_\pi)$ (see Appendix~\ref{app:winding} for the detailed derivation and numerical implementation). As seen in the phase diagram, starting from the QAH phase, increasing the $d$-wave pairing amplitude $\Delta$ continuously drives the system into this FMEB regime.

\section{Transport signatures in two-terminal and four-terminal geometries}\label{sec4}
From the edge spectra discussed in Sec.\ref{sec2}, the FMEB is characterized by two counterpropagating Majorana channels confined to the same boundary but separated in momentum space, in contrast to the single chiral edge channel of the QAH phase. Such coexistence of oppositely propagating Majorana modes suggests transport behavior fundamentally different from that of a chiral QAH insulator.

\begin{figure}[htbp]
    \centering
    \includegraphics[width=0.5\textwidth]{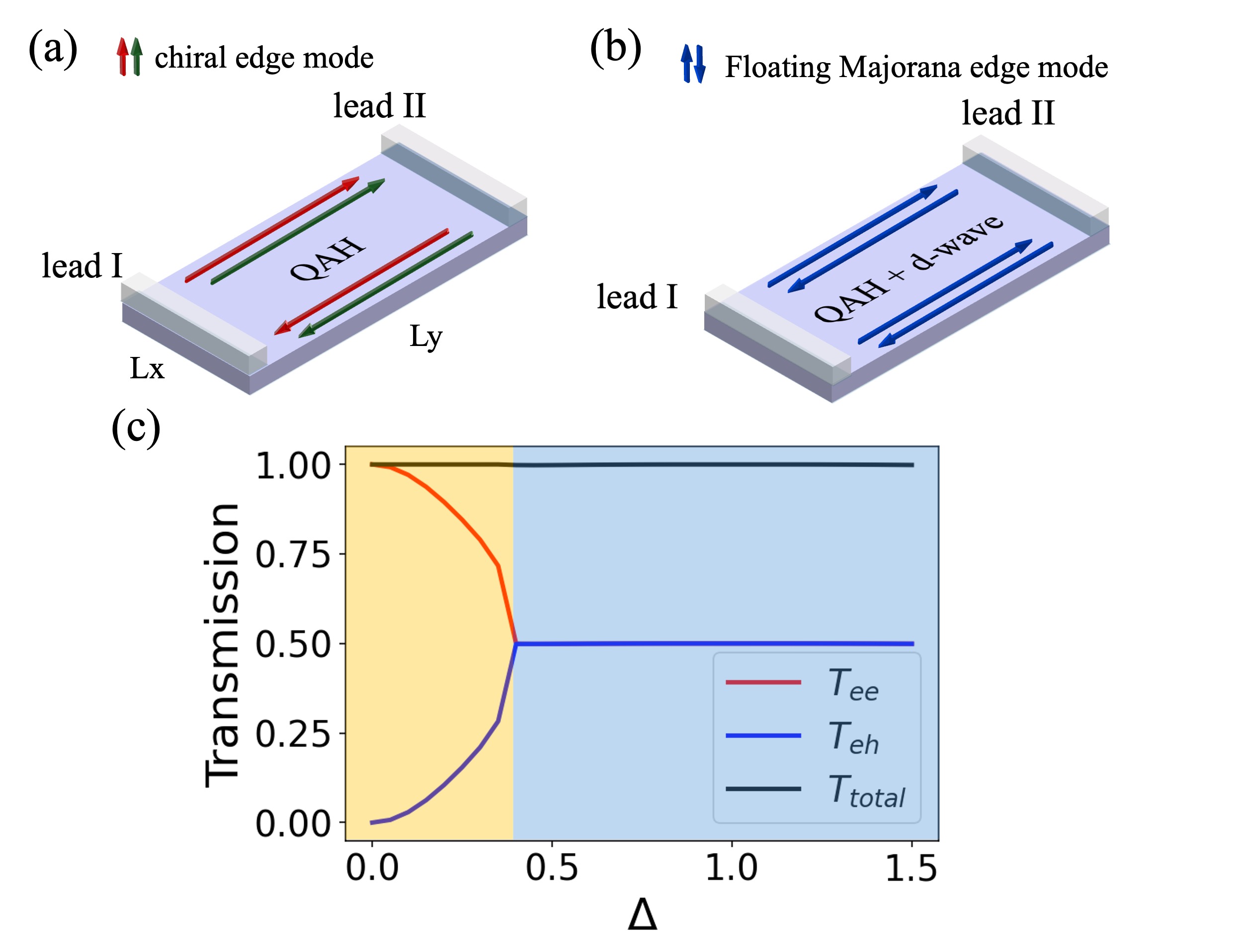}
        \caption{Two-terminal transport signatures across the QAH-FMEB transition. (a) In the absence of pairing, the system is in a chiral QAH phase with $\mathcal{N}=-2$, supporting a single chiral edge mode composed of co-propagating electron (red) and hole (green) channels. (b) Upon increasing the $d$-wave pairing strength, the edge reconstructs into a pair of counterpropagating Majorana modes forming an FMEB (blue). (c) Transmission coefficients as a function of $\Delta$: the total transmission $T_{\rm tot}\equiv T_{ee} + T_{eh}$ remains unity across the transition, while the components evolve from $(T_{ee}, T_{eh})\approx (1,0)$ in the QAH regime to $(1/2,1/2)$ in the FMEB regime, reflecting equal-weighted Majorana transport. Background shading indicates phase: yellow for QAH and blue for FMEB. System parameters: $m=-2.5, B=1$, ribbon size $L_x=L_y=80$.}
        \label{fig3}
\end{figure}

To elucidate these effects, we now examine the two-terminal transport response across the QAH $\rightarrow$ FMEB transition driven by the $d$-wave pairing amplitude $\Delta$, as summarized in Fig.\ref{fig3}.

Fig.\hyperref[fig3]{3(a)} shows the QAH ribbon without superconductivity, where the edge supports a single chiral edge mode composed of co-propagating electron and hole channels. In the BdG representation, the red and green lines represent electron and hole channels that together form a chiral edge mode with total Chern number $\mathcal{N}=-2$.
When the $d$-wave pairing becomes sufficiently strong, the resulting Wilson mass anisotropy drives the system into the FMEB regime, giving rise to a pair of counterpropagating Majorana channels, as shown by the blue line in Fig.\hyperref[fig3]{3(b)}.

To investigate the transport characteristics of the FMEB phase, we construct a two-terminal setup, as illustrated in Figs.~\hyperref[fig3]{3(a)} and \hyperref[fig3]{3(b)}. We discretize Eq.(\ref{eq:BdG_4band}) on a square lattice, with geometry size $L_x=L_y=80$. We employ the NEGF approach in the Nambu basis \((e,h)\), which allows us to compute the energy-resolved transmission coefficients\cite{PhysRevB.100.235407,mag,alter,li2024emergent,wan_quantum_2025}. For energy \(E\), the retarded Green’s function is \(G^{r}(E)=[E+i0^{+}-H_{\rm cen}-\sum_{p}\Sigma^{r}_{p}(E)]^{-1}\) with \(G^{a}=(G^{r})^{\dagger}\). Metallic leads are treated in the wide-band limit, \(\Sigma^{r}_{p}\simeq -\tfrac{i}{2}\Gamma_{p}\) and \(\Gamma_{p}=i(\Sigma^{r}_{p}-\Sigma^{r\dagger}_{p})\). The transmission from lead \(q\) to \(p\) can be obtained by
\begin{equation}
\mathbf{T}_{q\rightarrow p}(E)=\mathrm{Tr}\!\left[\Gamma_{p}G^{r}(E)\Gamma_{q}G^{a}(E)\right].
\label{eq:caroli}
\end{equation}
In Nambu representation, we decompose it into the electron-tunneling and crossed-Andreev reflection (CAR) parts,
\(\mathbf{T}^{ee}_{q\rightarrow p}(E)=\mathrm{Tr}[\Gamma^{ee}_{p}G^{r}_{ee}\Gamma^{ee}_{q}G^{a}_{ee}]\) and
\(\mathbf{T}^{eh}_{q\rightarrow p}(E)=\mathrm{Tr}[\Gamma^{hh}_{p}G^{r}_{he}\Gamma^{ee}_{q}G^{a}_{eh}]\).
We define the total transmission as $\mathbf{T}_{\mathrm{tot}}(E) = \mathbf{T}^{ee}_{I\rightarrow II}(E) + \mathbf{T}^{eh}_{I\rightarrow II}(E)$, since it directly determines the thermal conductance (see Appendix.\ref{APPENDIX A: Electric thermal conductance}). The left (lead~I) and right (lead~II) electrodes are attached to the opposite edges along the $x$-direction, each coupled over the full transverse width $L_y$. Lead-I and II are implemented as self-energy electrodes within the wide-band approximation.

Fig.\hyperref[fig3]{3(c)} summarizes the two-terminal transport response as the $d$-wave pairing strength $\Delta$ increases.
The yellow and blue regions correspond to the QAH and FMEB phases, respectively.
Throughout the entire $\Delta$ range, the total transmission
$\mathbf{T}_{\text {tot }}(E)=\mathbf{T}_{e e}(E)+\mathbf{T}_{e h}(E)$
remains close to unity, indicating that the effective number of conducting channels is conserved.
However, the composition of $\mathbf{T}_{\text {tot }}$ evolves qualitatively between the two phases.
In the QAH regime (yellow region), transport from lead~I to lead~II is dominated by a single chiral electron channel, as illustrated by the red line in Fig.\hyperref[fig3]{3(a)}. At $\Delta=0, \mathbf{T}_{e e} \simeq 1$ and $\mathbf{T}_{e h} \simeq 0$.
As $\Delta$ increases but remains below the critical value, electron and hole contributions gradually mix, leading to a partial conversion of the chiral electron channel into a charge-neutral superposition, reflected by the growth of $\mathbf{T}_{e h}$ and the concurrent decrease of $\mathbf{T}_{e e}$.
Beyond the transition, in the FMEB regime (blue region), the edge reconstructs into two counterpropagating Majorana modes localized on the same boundary but separated in momentum space.
Each Majorana mode carries half the transmission of a normal fermionic channel\cite{PhysRevB.105.125414}, resulting in an equal partition $\mathbf{T}_{ee}=\mathbf{T}_{eh}=1 / 2$ while maintaining the total $\mathbf{T}_{\text {tot }}=1$.
This invariance of $\mathbf{T}_{\text {tot }}$ thus reflects that the number of conducting channels is preserved across the QAH-FMEB transition, even though their microscopic nature changes from chiral electron transport to counterpropagating Majorana transport.

\begin{figure}[htbp]
    \centering
    \includegraphics[width=0.5\textwidth]{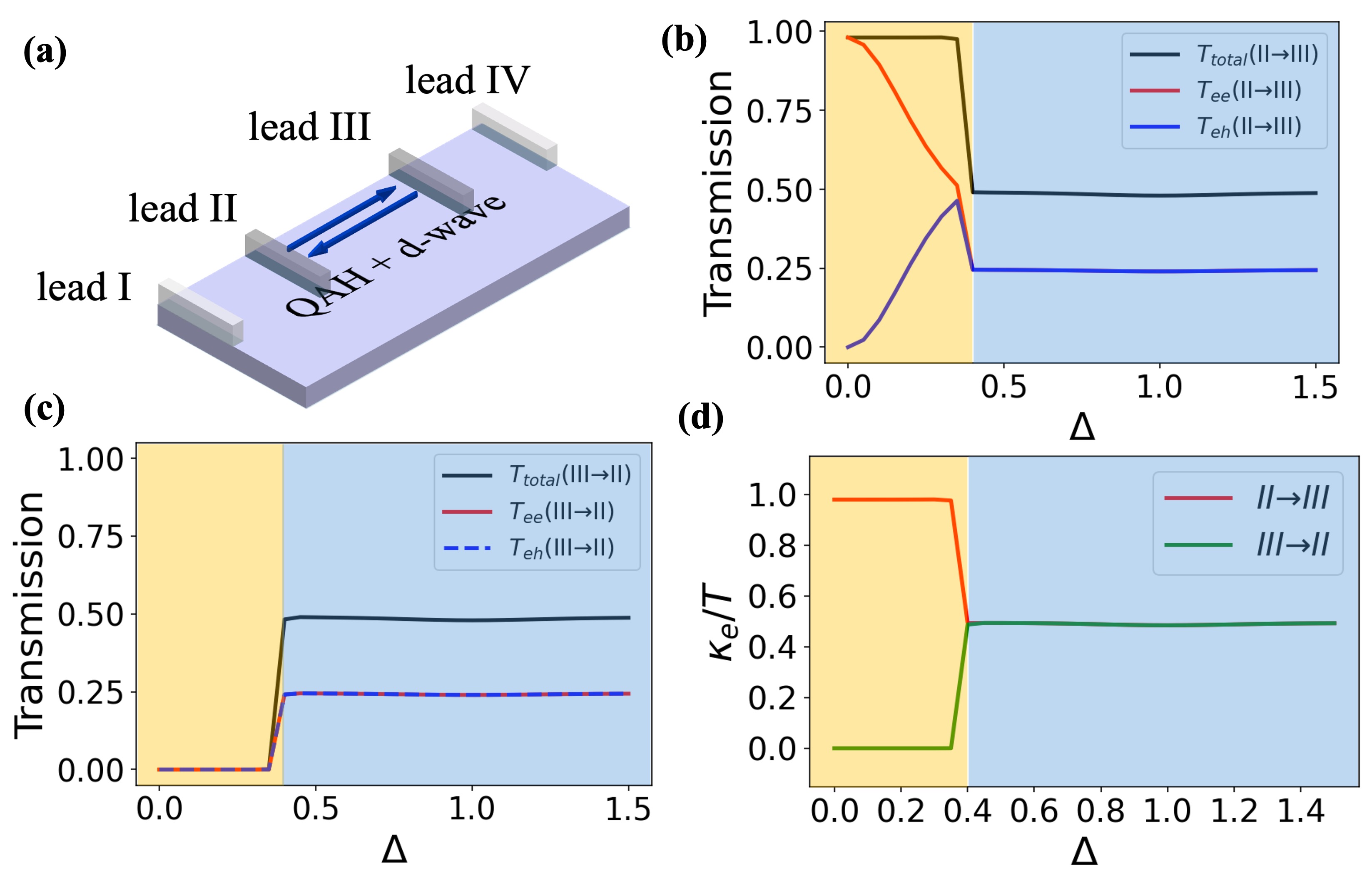}
    \caption{Four-terminal transport response distinguishing the QAH and FMEB phases. (a) Schematic of the four-terminal setup, where temperature-measuring leads II and III are locally coupled to the same sample edge, and leads I and IV act as floating terminals to suppress backscattering. (b,c) Direction-resolved transmissions as a function of pairing strength $\Delta$: in the QAH phase (yellow background), transport is chiral, with $\mathbf{T}_{\text {tot}}(II\to III)\approx 1$ and $\mathbf{T}_{\text {tot }}(III\to II )\approx 0$; in the FMEB phase (blue background), both directions yield $\mathbf{T}_{ee}\!\simeq\!\mathbf{T}_{eh}\!\simeq\!0.25$, reflecting bidirectional Majorana transport, with each mode contributing half of a normal channel. (d) Low-temperature thermal conductance $\kappa_e/T$ exhibits a robust half-quantized plateau in the FMEB regime, distinguishing it from the quantized chiral response in the QAH phase. Background color indicates phase: yellow for QAH and blue for FMEB. System parameters: $m=-2.5$, $B=1$, ribbon size $L_x=L_y=80$.}
        \label{fig4}
\end{figure}

The two-terminal measurement includes contributions from both the upper and lower boundaries, so the total number of conducting channels remains the same in the QAH and FMEB phases.
This explains why the total transmission $\mathbf{T}_{\text {tot }}$ is identical for the two phases despite their distinct edge structures.
To further distinguish them, we employ a four-terminal configuration in Fig.\hyperref[fig4]{4(a)}, where temperature-measuring leads (II and III) couple locally to a single edge, and protect leads (I and IV) suppress end reflections.

As shown in Fig.\hyperref[fig4]{4(b)}, the transmission from
lead~II to lead~III decreases from $\mathbf{T}_{\mathrm{tot}}=1$ to $\mathbf{T}_{\mathrm{tot}}=1/2$ as
$\Delta$ increases. In contrast to the two-terminal setup, the four-terminal geometry exposes the distinction between the QAH and FMEB phases through the behavior of $\mathbf{T}_{\mathrm{tot}}$ along the edge.
Concomitantly, $\mathbf{T}_{ee}$ falls from $1$ and locks to $1/4$ near the phase transition critical point, while $T_{eh}$ rises from $0$ and
jumps to $1/4$. The halving of
$\mathbf{T}_{\mathrm{tot}}$ reflects the reconstruction of a single chiral electron
channel in the QAH phase into a floating Majorana channel in the FMEB regime. A Majorana mode contributes effectively half a conventional fermionic channel and, being an equal electron-hole superposition, enforces
$\mathbf{T}_{ee}=\mathbf{T}_{eh}=1/4$ per direction. In the opposite direction, as shown in Fig.\hyperref[fig4]{4(c)}, the transmission from lead~III to lead~II is strongly suppressed in the chiral QAH phase (unidirectional edge transport) but rises to $\mathbf{T}_{\mathrm{tot}}=1/2$ upon entering the FMEB phase, with $\mathbf{T}_{ee}=\mathbf{T}_{eh}=1/4$, consistent with two counterpropagating Majorana modes coexisting on the same boundary at different momenta.

While the energy-resolved transmissions already expose the FMEB’s distinctive
composition, in practice, thermal conductance is the more accessible
observable in superconducting devices. The thermal response, evaluated in the low-temperature limit as $\frac{\kappa_e}{T}=\frac{\pi^2 k_B^2}{3 h}\left[\mathbf{T}_{e e}(E=0)+\mathbf{T}_{e h}(E=0)\right]$ (see Appendix.\ref{APPENDIX A: Electric thermal conductance}), is shown in Fig.\hyperref[fig4]{4(d)} (in units of $\pi^{2}k_{\rm B}^{2}/3h$).
In the QAH phase, the thermal conductance from lead~II to lead~III is quantized at $\pi^2 k_B^2 / 3 h$, while the reverse conductance (III $\rightarrow$ II) is nearly zero, reflecting the chiral nature of edge transport.
After the transition to the FMEB regime, both propagation directions along the same edge exhibit a robust half-quantized plateau, $\kappa_e / T=\pi^2 k_B^2 / 6 h$.
This evolution (from unidirectional quantized transport in the QAH phase to bidirectional half-quantized transport in the FMEB phase) provides a clear signature distinguishing the two regimes.
The symmetric half-quantized thermal conductance on a single edge is a characteristic hallmark of the FMEB and offers a concrete target for experimental verification.

Finally, we note that the corresponding two-terminal and four-terminal transport results for the $y$ direction are fully consistent with those shown here for the $x$ direction; the only difference is that the conducting FMEB channels originate from the other decoupled Majorana block due to the opposite $d$-wave-induced anisotropy in $H_{+}$ and $H_{-}$.

\section{Robustness of FMEB transport against disorder, temperature, and chemical potential}\label{sec5}
\subsection{Effect of disorder}\label{sec5a}
\begin{figure}[htbp]
    \centering
    \includegraphics[width=0.5\textwidth]{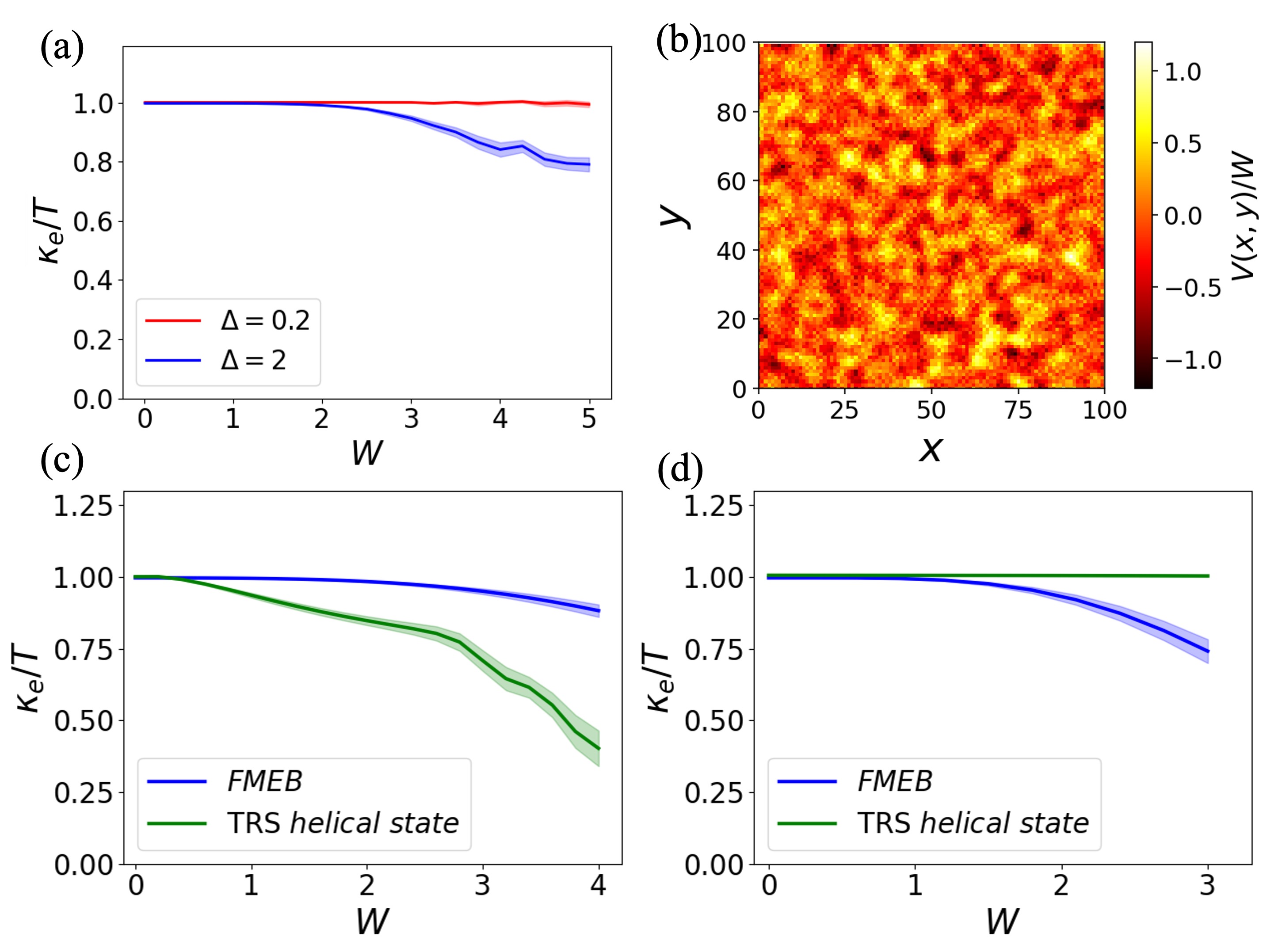}
        \caption{Robustness of FMEB thermal conductance against disorder. (a) Disorder-averaged thermal conductance $\kappa_e/T$ as a function of long-range correlated disorder strength $W$ for the chiral QAH phase (red, $\Delta=0.2$) and the FMEB phase (blue, $\Delta=2$). 
        (b) Representative disorder profile $V(x,y)/W$ used in the simulations, constructed via Gaussian filtering of white-noise disorder, with spatial correlation length $\eta\!\approx\!1.2a$. 
        (c) Disorder-averaged thermal conductance $\kappa_e/T$ 
        as a function of TRS-breaking long-range disorder strength $W$
        for the FMEB phase and the TRS helical Majorana edge state. 
        (d) Disorder-averaged thermal conductance $\kappa_e/T$ 
        as a function of Anderson-type potential disorder strength $W$. 
        Error bands in (a, c and d) indicate sample-to-sample standard deviation over 50 disorder realizations.
        System parameters: $m=-1$, $B=1$, $L_x=L_y=80$. The helical-edge model is given in Appendix \ref{The TRS helical Majorana edge model}. All conductance values are reported in units of $\pi^{2}k_{\rm B}^{2}/3h$.}
        \label{fig5}
\end{figure}

FMEBs host a pair of counterpropagating Majorana modes confined to the same boundary, closely resembling the helical Majorana edge states that appear in time-reversal-invariant topological superconductors (class~DIII). In contrast to the DIII case, however, the FMEB phase explicitly breaks time-reversal symmetry, and its overall Chern number vanishes ( $\mathcal{N}=0$ ). The coexistence of oppositely directed Majorana modes on a single edge without Kramers protection raises an important question regarding the stability of this phase. To address this issue, we next examine the robustness of FMEBs against long range disorder. We employ the same two-terminal setup as in Fig.\ref{fig2} (also used in Sec.\ref{sec5b} and Sec.\ref{ses5c}) to measure thermal transport and model long-range disorder by adding $V_i\,\tau_z\!\otimes\! \sigma_0$ to the device, with $V_i\equiv\epsilon_i$ constructed by Gaussian filtering of a white-noise source.
Here, $\epsilon_i$ is obtained by Gaussian smoothing of a uniform-Anderson disorder distribution\cite{PhysRevLett.103.156804,PhysRevB.83.235403,PhysRevB.97.235435,classify,new1}
\begin{equation}
\epsilon_i=\frac{\displaystyle\sum_{j}\tilde{\epsilon}_j\,
e^{-|\mathbf r_i-\mathbf r_j|^{2}/(2\eta^{2})}}
{\displaystyle\sqrt{\sum_{j}e^{-|\mathbf r_i-\mathbf r_j|^{2}/(2\eta^{2})}}},
\qquad \tilde{\epsilon}_j\in[-W/2,W/2],
\label{eq:LRdis_real}
\end{equation}
with $W$ the disorder strength and $\eta$ the correlation length (we take lattice constant $a=1$ and typically $\eta\simeq1.2\,a$). For each $W$, we average the transport results over 50 disorder realizations; shaded bands in Fig.\hyperref[fig5]{5(a)} indicate the sample-to-sample standard deviation, and a representative $V(x,y)$ is shown in Fig.\hyperref[fig5]{5(b)}.

Fig.\hyperref[fig5]{5(a)} compares $\kappa_e/T$ versus $W$ in the chiral QAH phase ($\Delta = 0.2$) and in the FMEB phase ($\Delta = 2$). With increasing disorder strength $W$, the QAH phase maintains a perfectly quantized thermal conductance throughout. In contrast, the FMEB phase exhibits a quantized plateau only for $W \lesssim 2$; upon further increasing $W$, the quantization gradually deteriorates and the plateau is eventually destroyed. The QAH edge's robust chiral propagation prevents backscattering, maintaining the quantized plateau $\kappa_e/T=1$ (in units of $\pi^{2}k_{\rm B}^{2}/3h$). By contrast, an FMEB consists of two counterpropagating Majorana channels.
Long-range disorder couples them only weakly because the oppositely propagating modes carry distinct momenta along the edge.
As a result, the half-quantized thermal plateau remains robust over a wide range of disorder strengths $W$.
When the disorder becomes sufficiently strong, however, large-momentum scattering processes emerge and mix the counterpropagating Majorana channels, leading to the breakdown of the quantized conductance. FMEBs tolerate long-range disorder due to the momentum separation of their counterpropagating edge modes. Nevertheless, the chiral QAH edge is even more resilient, as it lacks a counterpropagating partner for backscattering on the same boundary.

To further distinguish FMEBs from the conventional time-reversal-symmetric (TRS) helical Majorana edge states,
we perform two complementary disorder tests, as shown in Figs.\hyperref[fig5]{5(c)}  and \hyperref[fig5]{5(d)}.
In Fig.\hyperref[fig5]{5(c)}, we consider a long-range TRS-breaking disorder potential $V_i \, \tau_z\otimes \sigma_y$. In Fig.\hyperref[fig5]{5(d)}, we study an Anderson-type onsite potential disorder $w_i \, \tau_z\otimes \sigma_0 $
with $w_i$ uniformly distributed in $[-W/2,\, W/2]$, where $W$ characterizes the disorder strength.
The TRS helical Majorana edge model based on a $p$-wave superconductor used for comparison is provided in Appendix~\ref{The TRS helical Majorana edge model}. As shown in Fig.\hyperref[fig5]{5(c)}, for the TRS-breaking long-range disorder potential, the thermal conductance of the FMEB phase, $\kappa_e/T$, remains essentially quantized as the disorder strength increases, showing only a slight reduction when $W\gtrsim 2$. In sharp contrast, the quantized plateau of the TRS helical Majorana edge is highly sensitive to TRS breaking and is rapidly destroyed already at weak disorder.
As shown in Fig.\hyperref[fig5]{5(d)}, for the short-range Anderson-type potential disorder, the FMEB $\kappa_e/T$ is affected much more strongly and the quantized plateau gradually collapses once $W\gtrsim 1$. Meanwhile, the TRS helical Majorana edge retains an excellent quantized plateau under this TRS-preserving short-range disorder.
These contrasting disorder responses provide an operational distinction between FMEBs and TRS helical Majorana edge states.

\subsection{Temperature dependence}\label{sec5b}
\begin{figure}[htbp]
    \centering
    \includegraphics[width=0.32\textwidth]{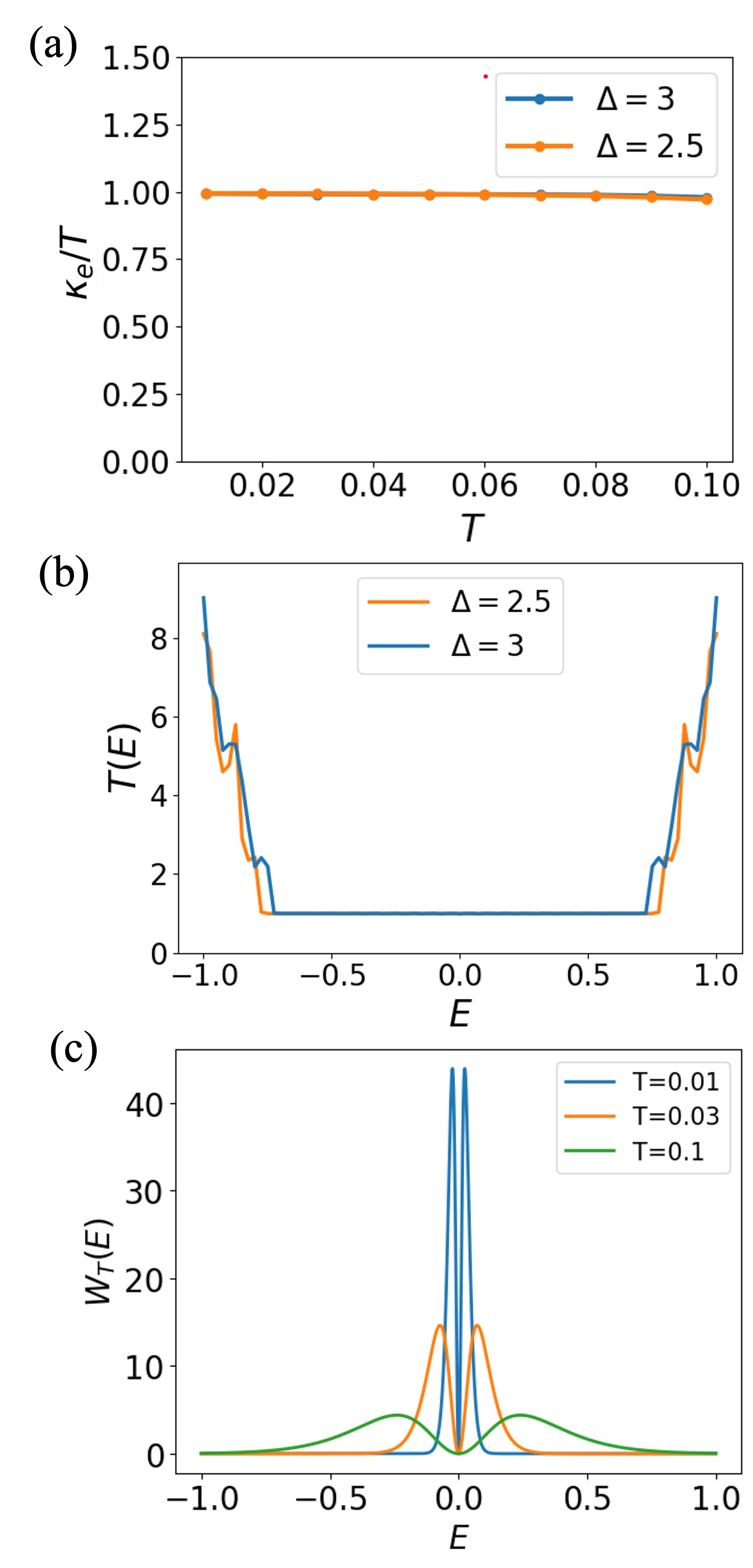}
     \caption{Temperature dependence of thermal transport in the FMEB regime. (a) Thermal conductance per unit temperature $\kappa_e/T$ versus temperature, shown for $\Delta = 2.5$ and $3$. In both cases, $\kappa_e/T$ remains close to unity across the full $T$ range, indicating robust quantization. (b) Energy-resolved total transmission $\mathbf{T}_{\rm tot}(E)$ shows a wide, nearly flat unit plateau around $E=0$, bounded by the gap edges. (c) Thermal weighting kernel $W_T(E)$ for $T=0.01, 0.03, 0.1$, showing an even function peaked at $E=0$; higher temperatures broaden the weighting but preserve normalization. System parameters: $m=-1$, $B=1$, $L_x=L_y=80$.}
        \label{fig6}
\end{figure}

All calculations discussed above have been performed within the zero-temperature approximation. We next investigate how finite temperature affects the quantized transport characteristics of the FMEB phase.
Within the Landauer–Büttiker framework (Eq.\ref{eq:kappa}), the thermal conductance can be written (in natural units) as
$\frac{\kappa_e}{T}
= \int_{-\infty}^{+\infty} dE\, W_T(E)\,\mathbf{T}_{\mathrm{tot}}(E)$, where
$W_T(E)=\frac{(E)^2}{(T)^3}\,f_0(E)\,[1-f_0(E)]$.
So that the temperature dependence enters exclusively through the thermal kernel \(W_T(E)\).
Fig.\hyperref[fig6]{6(a)} shows that in the FMEB regime, \(\kappa_e/T\) is essentially temperature independent, with only minor deviations when \(k_B T\) approaches \(0.1\).
This insensitivity signals a robust quantized response.
The transmission spectra in Fig.\hyperref[fig6]{6(b)} reveal a wide, nearly flat plateau with \(\mathbf{T}_{\mathrm{tot}}(E)\approx 1\) extending over \(|E|\lesssim 0.8\) for both \(\Delta=2.5\) and \(3\), ensuring that the thermal window samples a constant transmission.
As illustrated in Fig.\hyperref[fig6]{6(c)}, \(W_T(E)\) is an even function sharply peaked near \(E=0\) and rapidly decaying away from it; increasing \(T\) broadens and smoothens the peak but does not alter its normalization.
Consequently, as long as the thermal window defined by \(W_T(E)\) is contained within the flat transmission plateau, \(\mathbf{T}_{\mathrm{tot}}(E)\) may be treated as a constant under the integral, and using
$\int_{-\infty}^{+\infty} W_T(E)\,dE=1/3$ (in natural units)
one obtains a quantized $\kappa_e/T$.
These results demonstrate that the FMEB thermal conductance remains quantized within a certain temperature range, supporting its experimental observability.

\subsection{Finite chemical potential}\label{ses5c}

Experiments can tune the chemical potential ($\mu$) via gating or weak doping, yet achieving the exact charge-neutral point is generally difficult. To assess the relevance of such realistic conditions, we now examine how a finite $\mu$ impacts the FMEB and its thermal‐transport fingerprints.

In the Majorana representation, a small chemical potential couples the two Majorana blocks, leading to the following Hamiltonian:
\begin{equation}
\tilde{H}(\bm{k})
=
\begin{pmatrix}
H_+(\bm{k})&\mu\sigma_x \\\mu\sigma_x
 & H_-(\bm{k})
\end{pmatrix}.
\label{eq:H_tilde_lat}
\end{equation}

\begin{figure}[htbp]
    \centering
    \includegraphics[width=0.49\textwidth]{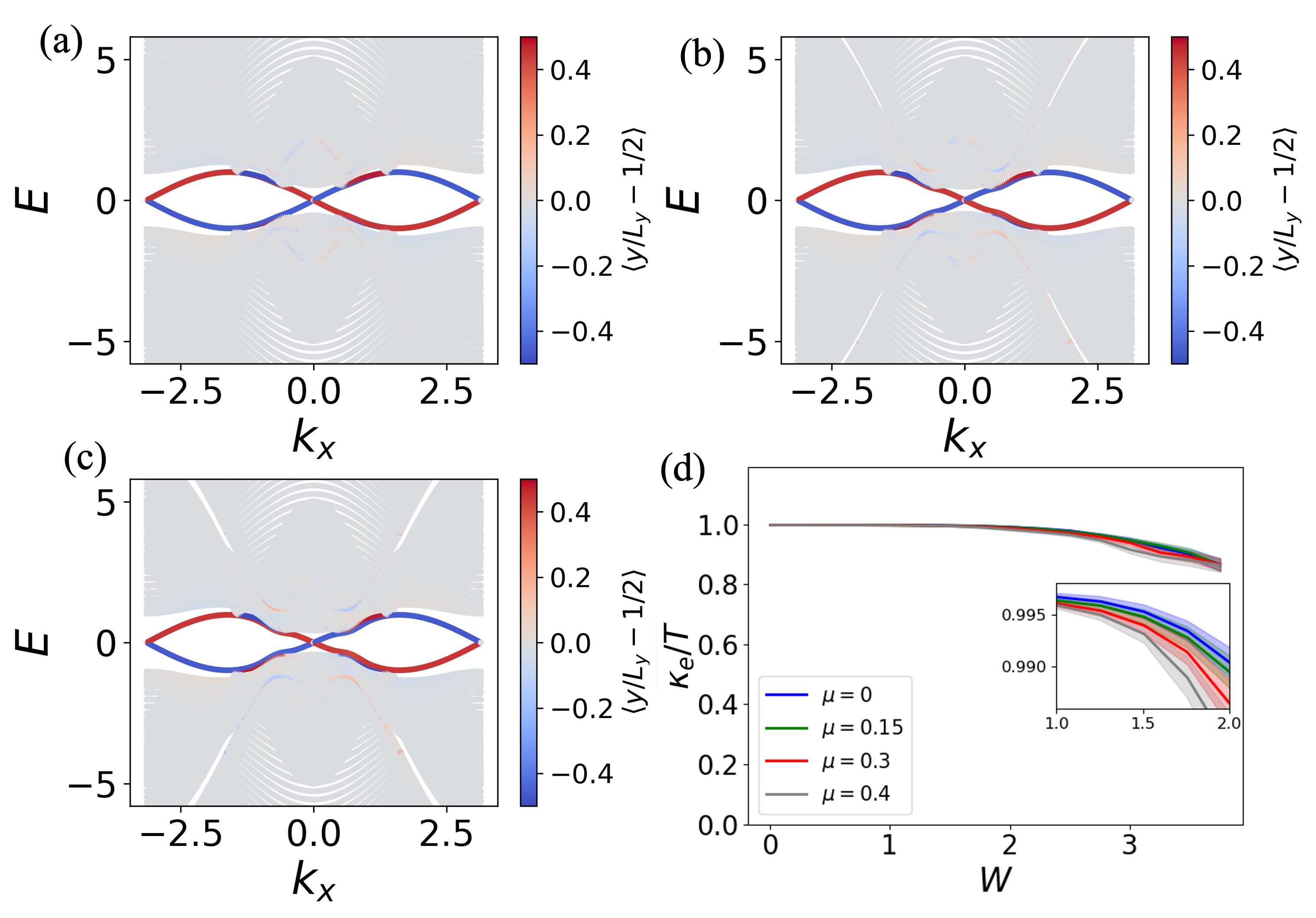}
        \caption{Effect of finite chemical potential on FMEB thermal transport. (a-c) Band structures of FMEB ribbons for increasing chemical potential $\mu=0.15, 0.30,0.40$, showing that the pair of counterpropagating Majorana edge modes remains intact, with only mild distortion in energy and momentum. Color indicates transverse localization $\langle y/L_y-1/2\rangle$, distinguishing upper versus lower edge states. (d) Thermal conductance $\kappa_e/T$ versus disorder strength $W$ for several values of $\mu$, showing nearly identical robustness across all cases. Conductance remains close to the quantized value $\pi^{2}k_{\rm B}^{2}/3h$ up to $W\approx 2$, indicating that FMEB transport is insensitive to moderate shifts in chemical potential. Inset shows a zoom-in of the low-disorder region, confirming minimal $\mu$-dependence. System parameters: $m=-1$, $B=1$, $L_x=L_y=80$. All conductance values are given in units of $\pi^{2}k_{\rm B}^{2}/3h$.}
        \label{fig7}
\end{figure}

Increasing $\mu$ (Fig.\hyperref[fig7]{7(a–c)} for $\mu = 0.15,0.30,0.40$) mildly distorts the helical FMEB pair and reduces the bulk gap, while the states remain well localized at the boundary (color scale).
Fig.\hyperref[fig7]{7(d)} shows the disorder dependence of the normalized thermal conductance $\kappa_e/T$  for several chemical potentials. For weak to moderate disorder $W\lesssim 2$, $\kappa_e/T$ remains pinned to its quantized value for all $\mu$. For stronger disorder, $W\gtrsim 2$, the conductance decreases monotonically and the reduction is more pronounced at larger $\mu$ (see the inset). Our results demonstrate the robustness of FMEB at small chemical potentials.

Overall, we have systematically assessed the stability of FMEB transport in two-terminal systems against realistic perturbations, including long-range disorder, finite temperature, and moderate chemical potential.
In the FMEB regime, the thermal conductance remains quantized up to disorder strengths of $W\lesssim 2$, owing to the momentum separation between counterpropagating Majorana modes, which suppresses backscattering. Thermal robustness arises from a broad, flat transmission plateau around $E=0$, ensuring insensitivity to thermal broadening. A finite chemical potential $\mu$ hybridizes the two Majorana blocks and slightly distorts the floating dispersion without destroying edge localization and the zero-energy transmission plateau. Consequently, the FMEB quantization of $\kappa_e/T$ is insensitive to small~$\mu$. Altogether, these results demonstrate that the FMEB is intrinsically robust against a wide class of realistic imperfections.

\section{Conclusions}\label{sec6}

In conclusion, we have identified FMEBs as a novel boundary phase in 2D topological superconductors that break time-reversal symmetry. Although the bulk Chern number vanishes, the edge hosts a momentum-separated pair of counterpropagating Majorana modes, forming an isolated, robust floating band. We demonstrated that this phase arises naturally in a QAH insulator proximitized by a $d$-wave superconductor, where pairing anisotropy generates the necessary Wilson mass imbalance in a microscopic, symmetry-consistent manner. Our NEGF analysis reveals that FMEBs exhibit clear thermal transport fingerprints: a preserved quantized conductance in two-terminal setups and a distinctive half-quantized plateau ($\kappa / T=\pi^2 k_B^2 / 6 h$) in single-edge four-terminal geometries. Importantly, this signature remains robust against finite temperature, long-range disorder, and moderate chemical potential, establishing FMEBs as a practically accessible phase. These results open a new route to realizing helical-like Majorana transport without time-reversal symmetry, and position FMEBs as a promising platform for exploring topological phases, edge reconstruction, and potentially fault-tolerant quantum computation.
Although our work is purely theoretical, it is conceivable that the underlying ingredients may be sought in a QAH thin film interfaced with a superconductor.
Candidate QAH platforms include magnetically doped (Bi,Sb)$_2$Te$_3$ and intrinsic magnetic topological insulators such as MnBi$_2$Te$_4$\cite{Chang2013Science_QAH,deng2020quantum}.
On the superconducting side, proximity from a cuprate ($d$-wave) superconductor into layered topological insulators has been demonstrated (e.g., via mechanical bonding) \cite{zareapour2012proximity}, and induced superconducting correlations in QAH-insulator devices have also been reported\cite{uday2024induced}.
We emphasize that, because cuprates are nodal, the induced pairing can retain a $d$-wave anisotropic component while its symmetry composition and effective momentum dependence may be sensitive to interface orientation and microscopic tunneling details\cite{PhysRevB.87.220506,PhysRevB.93.035140}.

\section*{ACKNOWLEDGMENTS}
This work was supported by National Natural Science Foundation of China (Grant Nos. 12474057, 12104028, 124B2069 and 12374034),
the National Key R and D Program of China (Grant No. 2024YFA1409002),
the Quantum Science and Technology-National Science and Technology Major Project (Grant No. 2021ZD0302403),
and the Fundamental Research Funds for the Central Universities.
The computational resources are supported by the High-Performance
Computing Platform of Peking University.

\appendix

\section{Winding number diagnosis of the FMEB phase}
\label{app:winding}

FMEBs are weak topology in the sense that they are tied to topological properties of quasi-1D subsystems at high-symmetry momenta, rather than to the 2D Chern number alone.
Although the FMEB phase share the same total
Chern number with a trivial phase ($\mathcal{N}=0$), they can be distinguished by an integer winding number defined for the quasi-1D subsystems at high-symmetry cuts.
In the generalized AZ classification\cite{PhysRevX.14.011057,hjs9-trrs}, these quasi-1D subsystems at $k_{y}=0/\pi$ (or equivalently $k_{x}=0/\pi$ for the
orthogonal ribbon) fall into class AIII with a $\mathbb{Z}^\times$ classification, and
Wannier localizability enables boundary modes to detach from the bulk continuum.
Therefore, it is sufficient to use the winding number\cite{ryu2010topological}
to characterize detached FEB phases.

In an $x$-open ribbon geometry, translation symmetry along $y$ is preserved and $k_y$ remains a good quantum number.
Fixing $k_y$ reduces each Majorana block to an effective 1D Bloch Hamiltonian $H_s(k_x;k_y)$, where
$s=\pm 1$ labels the two decoupled blocks.
At the two high symmetry points $k_y=0$ and $k_y=\pi$, one has $\sin k_y=0$, so the $\sigma_y$ component vanishes.
Consequently, the reduced 1D Hamiltonian $H_s(k_x;k_y=0/\pi)$ acquires an emergent chiral symmetry
\begin{equation}
\{C,\ H_s(k_x;k_y=0/\pi)\}=0,\qquad C=\sigma_y .
\label{eq:A3_chiral}
\end{equation}
To make this structure explicit, we work in the eigenbasis of $C$ using the unitary transformation
\begin{equation}
U=\frac{1}{\sqrt{2}}
\begin{pmatrix}
1 & 1\\
i & -i
\end{pmatrix},
\qquad
U^{-1} C\, U=\sigma_z .
\label{eq:A3_U}
\end{equation}
In this basis, the Hamiltonian becomes off diagonal,
\begin{equation}
U^{-1} H_s(k_x;k_y)\, U\Big|_{k_y=0/\pi}=
\begin{pmatrix}
0 & q_s^*(k_x;k_y)\\
q_s(k_x;k_y) & 0
\end{pmatrix},
\label{eq:A4_offdiag_q}
\end{equation}
where the complex function $q_s$ takes the form
\begin{equation}
\begin{aligned}
q_s(k_x;k_y)&= m + 2B(2-\cos k_x-\cos k_y)
\\&+ s 2\Delta(\cos k_x-\cos k_y) + iv\sin k_x,
\\&\qquad (s=\pm 1,\ \ k_y=0,\pi) .    
\end{aligned}
\label{eq:A4_qs}
\end{equation}

Hence, the winding numbers at $k_y=0/\pi$ is obtained as the standard formula\cite{ryu2010topological},
\begin{equation}
W_{s,0/\pi}=\frac{i}{2\pi}\int_{-\pi}^{\pi} dk_x\,
\big[q_s(k_x)\big]^{-1}\,\partial_{k_x} q_s(k_x).
\label{eq:A5_winding}
\end{equation}

\begin{figure}[htbp]
    \centering
    \includegraphics[width=0.49\textwidth]{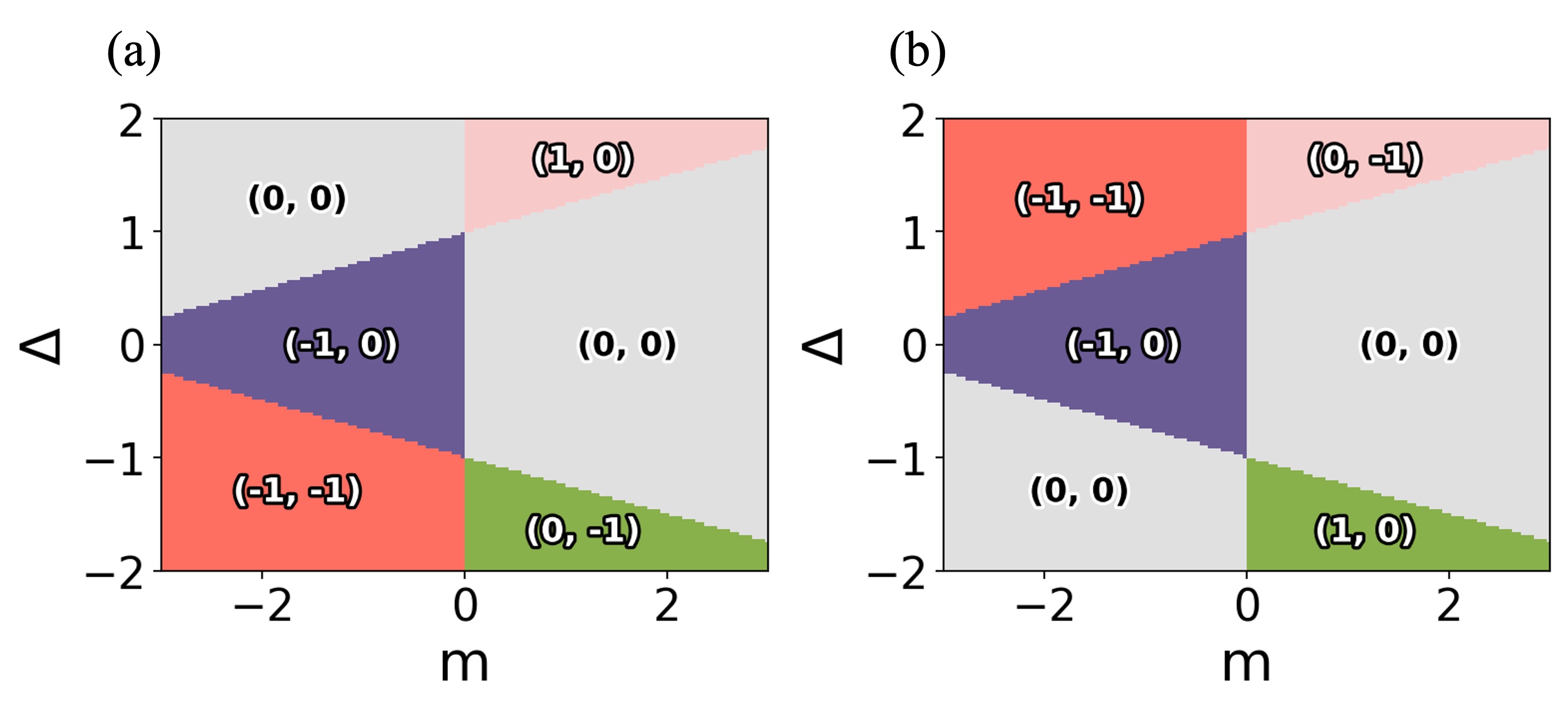}
        \caption{Winding number characterization of the two Majorana blocks.
(a) Phase diagram of $(W_0,W_\pi)$ for $H_{+}$ and (b) phase diagram of $(W_0,W_\pi)$ for $H_{-}$,
evaluated on the same parameter plane as Fig.\hyperref[fig2]{2(d)}.}

        \label{fig8}
\end{figure}

The $(W_{s,0},W_{s,\pi})$ thus provides a weak-type diagnostic that can distinguish phases even when the total Chern
number is unchanged. In particular, a truly trivial $N=0$ phase has $(W_{+,0},W_{+,\pi})=(0,0)$ and
$(W_{-,0},W_{-,\pi})=(0,0)$, whereas the FMEB regime features a nontrivial winding number in at least one block for a given
ribbon orientation $((W_{+,0},W_{+,\pi})=(\pm1,\pm1))$ or $(W_{-,0},W_{-,\pi})=(\pm1,\pm1))$. Fig.\hyperref[fig8]{8(a), (b)} show the winding number phase diagrams $(W_0,W_\pi)$ of the two Majorana blocks
$H_{+}$ and $H_{-}$, evaluated over the same parameter plane as in Fig.\hyperref[fig2]{2(d)}.
As shown in Fig.\hyperref[fig8]{8(a)}, the orange region corresponds to $(W_0,W_\pi)=(-1,-1)$ for $H_{+}$.
Since this region lies within the total Chern number sector $\mathcal{N}=0$ in Fig.\hyperref[fig2]{2(d)}, it is identified as an FMEB phase.
In contrast, the other $\mathcal{N}=0$ region in Fig.\hyperref[fig2]{2(d)} has $(W_0,W_\pi)=(0,0)$ for $H_{+}$ (Fig.\hyperref[fig8]{8(a)}),
and Fig.\hyperref[fig8]{8(b)} further shows that $H_{-}$ is also trivial there with $(W_0,W_\pi)=(0,0)$;
this region is therefore a trivial phase.
Meanwhile, the orange region in Fig.\hyperref[fig8]{8(b)} corresponds to $(W_0,W_\pi)=(-1,-1)$ for $H_{-}$,
which likewise occurs in the $\mathcal{N}=0$ sector and thus represents the other FMEB phase dominated by $H_{-}$.

\section{Electronic thermal conductance}\label{APPENDIX A: Electric thermal conductance}
Charge transport is suppressed by the charge neutrality of Bogoliubov quasiparticles (i.e., coherent superpositions of electrons and holes in the Nambu basis), whereas thermal transport remains a sensitive probe of Majorana physics. Indeed, quantized thermal conductance is widely regarded as a hallmark of Majorana edge modes.

After obtaining the transmission coefficients, the heat current under a small temperature bias
\((\mu_{L}=\mu_{R}=0,\; T_{L}=T+\Delta T,\; T_{R}=T-\Delta T)\)
can be expressed using the Landauer--B\"uttiker formula as
\begin{equation}
Q_{L}=\frac{1}{h}\!\int dE\, E\, \mathbf{T}_{\mathrm{tot}}(E)
\big[f(E,T_{L})-f(E,T_{R})\big],
\end{equation}
where \(f(E,T)=1/(e^{E/k_{B}T}+1)\) is the Fermi-Dirac distribution.

Expanding the Fermi functions to first order in the small temperature difference \(\Delta T\) yields the electronic thermal conductance\cite{PhysRevB.105.125414}:
\begin{equation}
\kappa_{e}=\frac{1}{h}\int_{-\infty}^{+\infty}\! dE\;
\frac{E^{2}}{k_{B}^{2}T^{2}}\,f_{0}(E)\,[1-f_{0}(E)]\,
\mathbf{T}_{\mathrm{tot}}(E),
\label{eq:kappa}
\end{equation}
with \(f_{0}(E)=f(E,T)\).

At relatively low temperatures, the electron contribution is restricted to a narrow energy window around the Fermi level, where \(\mathbf{T}^{ee}(E)\) and \(\mathbf{T}^{eh}(E)\) can be regarded as constant. In this limit, Eq.(\ref{eq:kappa}) reduces to the compact form
\begin{equation}
\frac{\kappa_{e}}{T} \simeq
\big[\mathbf{T}^{ee}(E=0)+\mathbf{T}^{eh}(E=0)\big]
\frac{\pi^{2}k_{B}^{2}}{3h}.
\end{equation}
Hence, the low-temperature thermal conductance depends only on the normal tunneling and crossed-Andreev reflection components evaluated at $E=0$.

Electronic thermal conductance exhibits a linear temperature dependence, whereas the phononic contribution scales cubically with \(T\). Consequently, at low temperatures, the electronic term dominates and provides a clear probe of Majorana-mediated heat transport.

\section{The TRS helical Majorana edge model}\label{The TRS helical Majorana edge model}

We summarize the TRS helical Majorana edge model used in Fig.\hyperref[fig5]{5}
as a system in class DIII, following the standard construction of a $p$-wave superconductor\cite{PhysRevLett.102.187001}. 
We start from a minimal 2D $p$-wave superconductor in class DIII. In the BdG form,
\begin{equation}
H_{\rm hel}=\frac12\int d^2x\;
\tilde{\Psi}^\dagger(\mathbf x)
\begin{pmatrix}
\varepsilon_{\mathbf p} & \Delta p_{+} & 0 & 0\\
\Delta p_{-} & -\varepsilon_{\mathbf p} & 0 & 0\\
0 & 0 & \varepsilon_{\mathbf p} & -\Delta p_{-}\\
0 & 0 & -\Delta p_{+} & -\varepsilon_{\mathbf p}
\end{pmatrix}
\tilde{\Psi}(\mathbf x),
\label{eq:helical_continuum_app}
\end{equation}
where $\tilde{\Psi}(\mathbf x)=\big[c_\uparrow(\mathbf x),c_\uparrow^\dagger(\mathbf x),
c_\downarrow(\mathbf x),c_\downarrow^\dagger(\mathbf x)\big]^T$,
$p_{\pm}=p_x\pm i p_y$, and $\varepsilon_{\mathbf p}$ denotes the normal-state dispersion.
Eq.(\ref{eq:helical_continuum_app}) is block diagonal and makes it explicit that spin-up (down) electrons form
$p_x+i p_y$ ($p_x-i p_y$) Cooper pairs, respectively.The two spin sectors carry opposite chiral $p$-wave pairings, which restores time-reversal symmetry and results in a Kramers pair of counter-propagating Majorana edge modes.

For numerical transport calculations, we use a standard square-lattice regularization.
In the Nambu basis $\Psi_{\mathbf k}=(c_{\mathbf k\uparrow},c_{\mathbf k\downarrow},
c_{-\mathbf k\uparrow}^\dagger,c_{-\mathbf k\downarrow}^\dagger)^T$,
we denote by $\tau_i$ the Pauli matrices in particle--hole space and by $\sigma_i$ those in spin space.
The Bloch Hamiltonian is
\begin{equation}
H_{\rm hel}(\mathbf k)=
\epsilon(\mathbf k)\,\tau_z\otimes \sigma_0
+\Delta\left[(\sin k_x)\,\tau_x\otimes \sigma_z
+(\sin k_y)\,\tau_y\otimes \sigma_0\right],
\label{eq:helical_lattice_k_app}
\end{equation}
with the lattice dispersion
\begin{equation}
\epsilon(\mathbf k)=-\mu+2t\left(2-\cos k_x-\cos k_y\right).
\label{eq:epsilon_k_app}
\end{equation}

In the above Nambu basis, the lattice Hamiltonian $H_{\rm hel}(\mathbf k)$ preserves time-reversal symmetry and therefore realizes a conventional TRS helical Majorana edge in the topological regime. Throughout this work, for the helical edge benchmark shown in Fig.\hyperref[fig5]{5}, we set $t=\mu=\Delta=1.0$. 
Consequently, the helical edge thermal plateau remains robust against TRS-preserving potential disorder, while it is rapidly degraded once TRS is explicitly broken (see Fig.\hyperref[fig5]{5}).

\bibliography{main}

\end{document}